\documentstyle[pra,aps,epsfig]{revtex}
\begin{document}
\draft
\twocolumn[\hsize\textwidth\columnwidth\hsize\csname @twocolumnfalse\endcsname
\author{E. Solano$^{1,2}$, R. L. de Matos Filho$^{1}$, and N. Zagury$^{1}$}
\title{Deterministic Bell states and measurement of the motional state of two
trapped ions }
\address{$^{1}$Instituto de F\'{\i}sica, Universidade Federal do Rio de Janeiro, Caixa Postal 68528, 21945-970 Rio de Janeiro, RJ, Brazil\\
$^{2}$Secci\'{o}n F\'{\i}sica, Departamento de Ciencias, Pontificia Universidad Cat\'{o}lica del Per\'{u}, Apartado 1761, Lima, Peru}
\date{December 2, 1998}
\maketitle

\begin{abstract}
  We present a method for the deterministic generation of all the
  electronic Bell states of two trapped ions. It involves the
  combination of a purely dispersive with a resonant laser excitation
  of  vibronic transitions of the ions.  In contrast to
  other methods presented up to now, our proposal does not require 
  differential laser addressing of the individual ions and 
  may be easily implemented with present available techniques. It is
  further shown that this excitation scheme is highly adequate for the
  complete determination of the motional state of the ions.
\end{abstract}

\vskip2pc]

The deterministic preparation of two-particle entangled states 
has become a subject of increasing interest. The quantum
correlations between the particles in such states may give rise to striking
phenomena, as the exclusion of a local realistic description of nature~\cite
{entangle}. Very recently it has been shown that, beside their interest in
studying the fundamentals of quantum physics, the deterministic generation
of two-particle entangled states is of extreme importance to experimentalists for the
implementation of quantum teleportation~\cite{telep}, quantum cryptography~\cite{crypto}  and quantum computation~\cite{qc}. Controlled 
entanglement between two massive particles has been achieved
both in the case of atoms crossing a high~$Q$ cavity\cite{haroche} and in
the case of ions in a trap\cite{twoions}. In the preparation and
manipulation of entangled states of two separated particles it is very
important to have a system that allows a well controllable coherent
interaction between the particles and an effective protection of the entangled
states against perturbations of the environment. The system composed by ions
in a linear trap meets these requirements very well~\cite{king}. In such
systems the interplay of the Coulomb interaction between the ions and their
coherent interaction with laser fields can be explored to create entangled
states of the ions.

Up to now most proposals presented to generate deterministic entangled
states of trapped ions rely on the necessity of addressing individual ions
by laser light~\cite{cirac,bollinger,garda}. As far as we know present
experimental conditions are such that the equilibrium distance between the
trapped and cooled ions is shorter than the beam waist of the applied laser
fields~\cite{king}. Therefore, presently,  laser beams cannot be focused
exclusively on individual ions. Recently, the consequences of this
experimental constraint have been partially circumvented in  Ref.~\cite{twoions}, by
introducing a differential displacement of the individual ions from the trap
center. Due to the ion micro-motion, the coupling constants between the
individual ions and the laser field become different. This allowed the
authors of ref.~\cite{twoions} to prepare electronic entangled states of
two ions, which are good approximations of two  Bell states.

In this Rapid Communication we present a scheme that allows the deterministic
generation of all the electronic Bell states of two trapped ions without the
requirement of differential laser addressing of the ions. This method is also
highly adequate to perform a complete determination of their vibrational
quantum state, including the possible entanglement between the two
vibrational modes along their alignment direction.

Let us consider the situation in which two trapped ions of mass $M$ are
cooled down to very low temperatures and aligned along the ${z}$ direction,
their equilibrium position, at zero temperature, being $z_{10}=d/2$ and $
z_{20}=-d/2$. We denote by $\hat{Z}=(\widehat{z_{1}}+\widehat{z_{2}})/2$
and $\hat{z}=\widehat{z_{1}}-\widehat{z_{2}}$ the center of mass and
relative coordinate position operators. Now consider that the ions are
excited by two classical homogeneous laser fields, $\vec{E}_{I}=\vec{E}
_{0I}e^{i(\vec{q}_{I}.\vec{r}-\omega _{I}t+\phi _{I})}$ and $\vec{E}_{II}$ = 
$\vec{E}_{0II}e^{i(\vec{q}_{II}.\vec{r}-\omega _{II}t+\phi _{II})}$, with $
\vec{q}_I$ and $\vec{q}_{II}$ parallel to the $z$ direction, which are quasiresonant
with a long living electronic transition among two hyperfine levels $|\downarrow \rangle $ and $|\uparrow \rangle $ of the ions. In the
Schr\"{o}dinger picture the interaction Hamiltonian describing this
situation is given by 
\begin{eqnarray}
\widehat{H}=&&\sum_{\alpha =I,II}\hbar \Omega _{\alpha }\lbrack e^{i(q_{\alpha }\hat{Z}
-\omega _{\alpha }t+\phi _{\alpha })} \nonumber 
\\ 
&&\times (\widehat{D}_{1}e^{iq_{\alpha }\hat{z}
/2}+\widehat{D}_{2}e^{-iq_{\alpha }\hat{z}/2})+{\rm H}.c. \rbrack
\label{H1}
\end{eqnarray}
where $\Omega _{\alpha }$ ($\alpha =I,II)$ are the Rabi classical
frequencies corresponding to the one photon transitions and
$\widehat{D}_{j}$ is the electric dipole transition operator
associated with the ion at the position $\hat{z} _{j}$ $(j=1,2)$. It
can be written as $\widehat{D}_j=\widehat{S}_{+j}+\widehat{S}_{-j}$,
with $\widehat{S}_{+j}=|\!\!\uparrow_j\rangle\langle
\downarrow_j\!\!|$ and
$\widehat{S}_{-j}=|\!\!\downarrow_j\rangle\langle \uparrow_j\!\!|$
being the raising and lowering operators associated with the two
electronic levels, $|\!\!\downarrow\rangle$ and $|\!\!\uparrow\rangle$,
 of frequencies $\omega _{\downarrow }$ and $\omega _{\uparrow
  }=\omega _{0}+\omega _{\downarrow }$, respectively (see Fig.~1).

In practice the situation described above could be effectively
realized by two pairs of laser fields so that each laser field
interaction in Eq.~(\ref{H1}) stands for a Raman interaction induced
by a pair of laser beams.  The lasers belonging to the first (second)
pair would have frequencies $\omega _{1}$ and $\omega _{2}$ ($\omega
_{1}^{\prime }$ and $\omega _{2}^{\prime }$) so that $\omega
_{I}=\omega _{1}-\omega _{2}$ ($\omega _{II}=\omega _{1}^{\prime
  }-\omega _{2}^{\prime }$). We notice that the relative phase of the
lasers in the first and in the second pair should correspond to $\phi
_{I}$ and $\phi _{II}$ in Eq.~(\ref{H1}), respectively. To
effectively generate the interaction described by Eq.~(\ref{H1}) these
two laser pairs should independently connect the electronic levels
$|\downarrow \rangle $ and $|\uparrow \rangle $ of both ions through a
virtual third level $|c\rangle $. This will be the case if the
difference $|$ $\Delta -\Delta ^{\prime }|$ in the
detunings, $\Delta =\omega _{c}-\omega _{\downarrow }-\omega _{1}$ and $
\Delta ^{\prime }=\omega _{c}-\omega _{\downarrow }-\omega _{1}^{\prime }$,
is much larger than the eigenfrequencies of the vibrational motion.

\vspace*{-3.0cm}
\begin{figure}
\begin{center}
\centerline{\leavevmode \epsfxsize=9cm\epsfbox{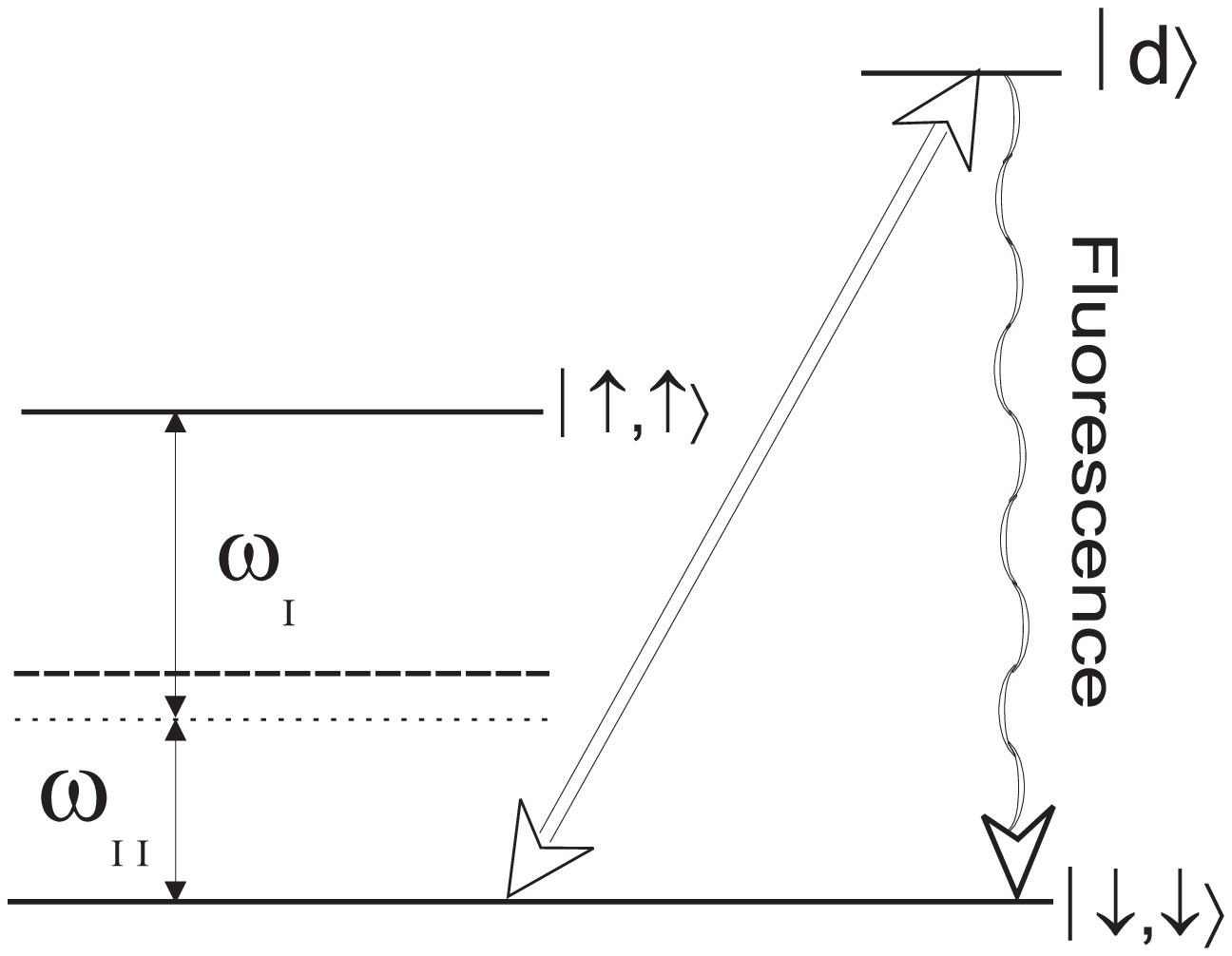}}

\vspace*{-4.0cm}
\caption{ Energy level diagram. A stimulated electronic transition is
  induced by means of two dispersive excitations with frequencies
  $\omega_{I}=\omega _{o}+k\nu-\delta$ and
  $\omega_{II}=\omega_{o}-k\nu+\delta$.  The detection of the
  electronic state, $|\downarrow ,\downarrow\rangle $ or $|\uparrow
  ,\uparrow \rangle $, is provided by the fluorescence resulting from
  optical pumping of the ions with a pulse tuned to the cyclic
  transition $|\downarrow ,\downarrow \rangle\Longleftrightarrow $
  $|d\rangle $. }
\end{center}
\label{fig1}
\end{figure}

From now on we will consider only the effective interaction described in Eq.~(\ref
{H1}), taking into consideration that $\phi _{I}$ and $\phi _{II}$ may be set
at will.  Furthermore, for simplicity, we take $\Omega _{I}=\Omega
_{II}=\Omega ,$ $q_{I}=q_{II}=q$ and $\phi _{I}=\phi _{II}=\phi .$ We
discuss a situation where the effective transitions described in Eq.~(\ref{H1})
are quasiresonant to one of the vibrational modes, say, the
center-of-mass one, which we take as having frequency $\nu $. Dispersive interactions
quasiresonant to a vibrational mode have already been considered in refs. 
\cite{nicimjc,garda}. Assuming that $
\omega _{I}=\omega _{0}+k\nu -\delta $ and $\omega _{II}=\omega
_{0}-k^{\prime }\nu +\delta ^{\prime },$ where $\delta ,\delta
^{\prime }\ll \nu $, and discarding the rapidly oscillating terms in
Eq.~(\ref{H1}), the  Hamiltonian in the interaction picture
may be written as
\begin{eqnarray}
\widehat{H} =&&\hbar \Omega \lbrack(\widehat{S}_{+1}e^{i\phi _{0}/2}+\widehat{S}_{+2}e^{-i\phi
_{0}/2})((i\eta )^{k}\hat{a}^{\dagger k}\widehat{F}_{k}( \hat{n}_c,\hat{n}
_{r}) e^{i\delta t}  \nonumber \\
&&+(i\eta )^{k^{\prime }}\widehat{F}_{k^{\prime}}
( \hat{n}_c,\hat{n}_{r}) \hat{a}^{k^{\prime }}e^{-i\delta
^{\prime }t})e^{i\phi }+{\rm H}.c.\rbrack
\label{H2}
\end{eqnarray}
\noindent
Here, $\phi _{0}=qd$ is the phase difference due to the spacing
between the ions, $\hat{a},\hat{b}$ and $\hat{a}^{\dagger
  },\hat{b}^{\dagger }$ are the annihilation and creation operators
associated with the excitations of the center of mass 
and  relative motion modes, respectively. The corresponding number operators are 
$\hat{n}_c=\hat{a}^{\dagger }\hat{a}$ and
$\hat{n}_{r}=\hat{b}^{\dagger }\hat{b}$ and  
\begin{equation}
\widehat{F}_{k}\left( \hat{n}_c,\hat{n}_{r}\right) =\sum
f_{k}(n_c, n_{r})|n_c,n_{r}\rangle \langle n_c,n_{r}|  \label{efer}
\end{equation}
with 
\begin{equation}
f_{k}(n_c,n_r)=e^{-(\eta ^{2}+\eta_r^{2})/2}\frac{n_c!}{(n_c+k)!}L_{n_c}^{k}(\eta ^{2})L_{n_r}^{0}(\eta _{r}^{2}).  \label{efek}
\end{equation}
$\eta =q\sqrt{\hbar /4M\nu }$ and $\eta _{r}=q\sqrt{\hbar /4M\nu _{r}}
$ are the Lamb-Dicke parameters associated to the center of mass and
relative vibrations, respectively, and $L_{m}^{k}(x)$ are associated
Laguerre polynomials.
For the Coulomb interaction among the two  ions it can be easily shown that $
\nu _{r}=\sqrt{3}\nu $, so that $\eta _{r}=\eta /\sqrt{4}{3}$.  When $k=k^{\prime }$ and 
$\delta ^{\prime }=\delta $,  so that $\omega
_{I}+\omega _{II}=2\omega _{0}$
\cite{steinbach}, 
Eq.~(\ref{H2}) reduces to the simpler
result 
\begin{eqnarray}
\widehat{H}&=&\hbar \Omega \lbrack (i\eta
)^{k}(\widehat{S}^{\prime}_{+1}+\widehat{S}^{\prime}_{+2})e^{i\phi }+{\rm H}.c. \rbrack  \nonumber \\
&&\times \lbrack \hat{a}^{\dagger k}\widehat{F}_{k}\left( \hat{n}_c,\hat{n}
_{r}\right) e^{i\delta t}+\widehat{F}_{k}\left( \hat{n}_c,\hat{n}_{r}\right) 
\hat{a}^{k}e^{-i\delta t}\rbrack,\label{Heff1}
\end{eqnarray}
where $\widehat{S}^{\prime}_{+1}=\widehat{S}_{+1}e^{i\phi_{o}/2}$ and
$\widehat{S}^{ \prime}_{+2}=\widehat{S}_{+2}e^{-i\phi _{o}/2}$. 
A similar result may be obtained if we choose to excite the stretch mode. Note that for $k \ne 0,1$  we may have $k\nu -\delta $ $=m\sqrt{3}\nu -\bar\delta$, with $\delta, \bar\delta \ll\nu$, $\eta_r^m/\bar{\delta} \approx O(\eta^k/{\delta})$ and both modes could be excited simultaneously. This will not be the case if we exchange the roles of the stretch and center of mass modes in Eq. (\ref{Heff1}). When $\delta $ is much larger than the vibronic Rabi frequency, $\eta
^{k}\Omega f_{k}(n_c, n_{r})$, and for interaction times $\delta t\gg
1,$ the Hilbert space of the electronic states effectively
decouples into two subspaces: one containing only one ion in the
excited state and the other containing both ions either in the excited or in
the ground state. In this case the effective  Hamiltonian
reads
\begin{eqnarray}
\widehat{H}_{\rm eff} &=&\hbar\Omega _{k}\lbrack \widehat{S}_{+1}^{\prime }\widehat{S}_{+2}^{\prime
}e^{2i\phi }+(-1)^k\left( \widehat{S}_{+1}^{\prime }\widehat{S}_{-2}^{\prime }+\frac{1}{2}\right)
\rbrack   \nonumber \\
&&\times \widehat{F}_{k}^{2}(\hat{n}_c,\hat{n}_{r})\lbrack \frac{\hat{n}_c!}{\left( \hat{n}_c-k\right) !}-\frac{\left( \hat{n}_c
+k\right) !}{\hat{n}_c!}\rbrack
+{\rm H}.c.  \label{Heff}
\end{eqnarray}
where $\Omega _{k}=2 |\Omega |^{2}\left( i\eta \right) ^{2k}/\delta$. The first term and its Hermitian conjugate describe two-photon
processes leading to the simultaneous excitation or deexcitation of
the electronic states of the two ions.  The second term and its
Hermitian conjugate describe processes where one ion undergoes a
transition from the ground to the excited electronic state and the
other ion makes a transition in the inverse direction, both processes
taking place simultaneously.  The third term is the contribution from
the interaction to the self energy of the ions. Note that processes
leading to independent excitation of the ions are not present anymore.

Assume that we have prepared the two-ion system in the state $\left|
\downarrow ,\downarrow \right\rangle _{n_c,n_{r}}\equiv \left| \downarrow
,\downarrow \right\rangle \otimes \left| n_c,n_{r}\right\rangle .$ Applying
the two laser pairs simultaneously, this initial state evolves after a time $t
$ to 
\begin{eqnarray}
e^{-\frac{i}{\hbar}\widehat{H}_{\rm eff}t}\left| \downarrow, \downarrow \right\rangle_{n_c,n_r} =&&
e^{-i(-1)^{k}\Omega _{n_cn_{r}}^{k}t} \lbrack \cos ( |\Omega
_{n_cn_{r}}^{k}|\,t) \left| \downarrow, \downarrow \right\rangle
\nonumber \\
&&+ i(-1)^{k}e^{2i\phi }\sin ( |\Omega _{n_cn_{r}}^{k}|\,t)
\left| \uparrow, \uparrow \right\rangle \rbrack \nonumber \\
&&\otimes\left| n_c,n_{r}\right\rangle,
\label{evol}
\end{eqnarray}
\noindent where 
\begin{equation}
\Omega _{n_c n_{r}}^{k}=\Omega_k f_{k}(n_c ,n_{r})^2 \lbrack \frac{n_c !}{(n_c -k)!}-\frac{
\left( n_c +k\right) !}{n_c !}\rbrack  \label{rabi}
\end{equation}
are effective Rabi frequencies for transitions in the Hilbert subspace
spanned by the states $\left| \downarrow ,\downarrow \right\rangle
_{n_c,n_{r}}$ and $\left| \uparrow ,\uparrow \right\rangle
_{n_c,n_{r}}.$ Note that, in this approximation, the interaction is
purely dispersive, so that it does not change the number of
vibrational excitations, leaving vibrational Fock states
$|n_c,n_r\rangle$ unchanged. Also, the two-ion state with one ion in
the excited state and the other in the ground state is never
populated, if the initial state is any combination of states where the
ions are both in their excited or ground electronic states. Therefore,
the time evolution described by Eq.~(\ref{evol}) corresponds to an
\lq \lq effective two-photon transition\rq \rq among the two states $\left|
  \downarrow ,\downarrow \,\right\rangle $ and $\left| \uparrow
  ,\uparrow \,\right\rangle $.  In this sense they could be considered
as a collective two-level system. This fact can be exploited for
measuring the vibrational state of the two-ion system, as will be shown
below.

Equation~(\ref{evol}) shows that we may generate  any combination of the ground, $
\left| \downarrow , \downarrow  \right\rangle$, and the doubly excited
states, $\left| \uparrow , \uparrow \right\rangle$, and, in particular, the two orthogonal Bell states  

\begin{equation}\label{phi}
|\Phi ^{(\pm )}\rangle =\frac{1}{\sqrt{2}}\lbrack \left| \downarrow
,\downarrow \right\rangle \pm i(-1)^{k}e^{2i\phi}\left| \uparrow
,\uparrow \right\rangle \rbrack\, , 
\end{equation}
if we let the lasers interact with the ions for either a time $t_{+}=\pi /(4|\Omega
_{n_cn_{r}}^{k}|)$ (to generate the state $|\Phi ^{(+)}\rangle $) or a time $
t_{-}=3\pi /(4|\Omega _{n_cn_{r}}^{k}|)$ (to generate the state $|\Phi
^{(-)}\rangle $). Alternatively, we could maintain the same time of
interaction and change the value
of  $\phi$ by $\pi/2 .$  

We will show now that it is possible to prepare the other two Bell states
applying a  laser pulse resonant to the electronic transition  
$|\uparrow\rangle\leftrightarrow |\downarrow\rangle $ (carrier pulse) 
 to one of the $|\Phi ^{(\pm )}\rangle $ states 
prepared with the dispersive interaction. For a carrier pulse  we just
need to deal with one effective laser $\vec{E}=\vec{E_{0}}e^{i(\vec{q_{0}}.
\vec{r}-\omega _{0}t+\varphi )}$, with $\vec{q_{0}}$ in the $z$ direction. 
In this case the effective Hamiltonian is
\begin{equation}
\widehat{H}=\hbar \Omega \lbrack (S_{+1}e^{i\varphi
_{0}/2}+S_{+2}e^{-i\varphi_{0}/2})e^{i\varphi }+{\rm H}.c.\rbrack \widehat{F}_{0}\left( \hat{n}_c,\hat{n}_{r}\right) ,
\label{Hcarrier}\end{equation}
where $\Omega $ is the Rabi frequency associated with the Raman  transition
and $\varphi _{0}=q_{0}d.$ 

After we have prepared one of the states $|\Phi ^{(\pm )}\rangle
|n_c,n_{r}\rangle $   we apply the  carrier pulse during a time $t_{0}.$ The time evolution of $|\Phi ^{(\pm
)}\rangle |n_c,n_{r}\rangle $ is then given by 

\begin{eqnarray}
&&\frac{1}{\sqrt{2}}\lbrace \lbrack \cos ^{2}\left( \Omega _{0}t_{0}\right) \mp \sin
^{2}\left( \Omega _{0}t_{0}\right) e^{-2i\varphi }\rbrack \left| \downarrow ,
\downarrow \right\rangle  \nonumber \\
&&-\frac{i}{2}\sin \left( 2\Omega _{0}t_0\right) 
(e^{i\varphi }\pm e^{-i\varphi })\left( e^{i\varphi _{0}/2}\left| \uparrow ,
\downarrow  \right\rangle +e^{-i\varphi _{0}/2}\left| \downarrow , \uparrow 
\right\rangle \right)  \nonumber \\
&&+ \lbrack \pm \cos ^{2}\left( \Omega _{0}t_0\right) -\sin ^{2}\left( \Omega
_{0}t_0\right) e^{2i\varphi }\rbrack \left| \uparrow , \uparrow  \right\rangle \rbrace
\otimes|n_c,n_{r}\rangle
\end{eqnarray}
where $\Omega _{0}=\Omega f_{0}\left( n_c,n_{r}\right) $. If we let
the pair of Raman lasers interact with the ions during
$t_{0}=\pi/(4|\Omega_{0}|)$
and choose their relative phases equal to $0$ (modulus $\pi )$  or  $
\varphi =\pi /2$ (modulus $\pi ),$ accordingly if we start with $|\Phi
^{(+)}\rangle $ or $|\Phi ^{(-)}\rangle , $ we generate the  Bell state 

\begin{equation}\label{psi}
\frac{1}{\sqrt{2}}\lbrack \left| \uparrow
,\downarrow \right\rangle +e^{i\varphi_0}\left| \uparrow
,\downarrow \right\rangle \rbrack\, . 
\end{equation}
The state orthogonal to this one may be generated by changing   $\varphi _{0}$ to  $\varphi _{0}+\pi,$ as was done in Ref.~\cite{twoions}.

It is remarkable that, in generating these Bell
states, we need neither to be precise about which sideband $k$ is  excited 
nor to claim the validity of the Lamb-Dicke approximation. This is a
consequence of the fact that the interactions we are considering do not
change the values of the vibrational quanta, either because they are purely
dispersive \lbrack Eq.~(\ref{Heff})\rbrack \hspace {2pt}or correspond to a carrier pulse \lbrack Eq.~(\ref{Hcarrier})\rbrack.

Until now we have considered the ions to be in a vibrational two-mode Fock state. The general state that describes a motional thermal state, when the ions are cooled down to a temperature close to
zero, is a density operator of the form 
\begin{equation}
\hat{\varrho}_v =\sum \Pi_{n_c,n_{r}}\,  \left|{n_c,n_{r}} \right\rangle
\langle {n_c,n_{r}}|\, ,
\end{equation}
where $\Pi_{n_c,n_{r}}=\left\langle n_c,n_{r}\right| \hat{\varrho}
_{v}\left| n_c,n_{r}\right\rangle $ is the vibrational population. As
$\Omega _{n_c n_{r}}^{k}$ depends on the values of $n_c$ and $n_{r}$
our proposal to generate the Bell states will not work in general. In
this case, there are several different contributions to the
probability of exciting the vibronic transitions, which may interfere
without control when the temperature departs from zero. 
Notice that thermal states at very low temperature, such that the probability of finding $n_r\approx 99\%$ and $n_c\approx 90\%$, has already been reported in ref.\cite{king}. 
On the other hand, if we
consider excitations in the first sideband ($k=1)$ and if the
Lamb-Dicke limit is valid \lbrack $f_k(n_c,n_r)\rightarrow 1$\rbrack, the effective
Rabi frequencies are independent of the vibrational state of the ions. This fact
has been noticed before by S{\o}rensen and M{\o}lmer\cite{garda}. It
allows the generation of electronic Bell states even if the ions are
in a thermal state with equilibrium mean vibrational numbers much less
than the inverse of the Lamb--Dicke parameters.

Let us now turn our attention to the determination of a general
vibrational quantum state, $\hat{\varrho}_{v}$, of the ions. Our
method is based on the determination of the two-mode Wigner function
$W(\alpha_1,\alpha _{2})$, characterizing the joint state of two
harmonic quantum oscillators at the point $(\alpha _{1},\alpha _{2})$
of their extended phase space, which can be written as~\cite{wal1}
\begin{equation}
W(\alpha _{1},\alpha _{2})=\frac{4}{\pi ^{2}}\sum_{n_{1},n_{2}}^{\infty
}(-1)^{n_{1}+n_{2}}\,\Pi_{n_1,n_{2}}(-\alpha _{1},-\alpha _{2})\, , 
\label{wigner}
\end{equation}
where $\Pi_{n_1,n_{2}}(-\alpha _{1},-\alpha _{2})$ is the population
corresponding to the two-mode harmonic oscillator state displaced
coherently in the extended phase space by $-\alpha _{1}$ and $-\alpha
_{2}$, respectively. To determine the vibrational Wigner function,
$W(\alpha_c, \alpha_r)$, using the above relation
\lbrack Eq.~(\ref{wigner})\rbrack, we first displace coherently the state
$\hat{\varrho}_{v}$ in the extended phase space
\begin{equation}
\hat{\varrho}_v(-\alpha _{c},-\alpha _{r})=\hat{D}_{c}^{\dagger }(\alpha _{c})
\hat{D}_{r}^{\dagger }(\alpha _{r})\hat{\varrho}_v\hat{D}_{r}(\alpha _{r})\hat{
D}_{c}(\alpha _{c}).
\end{equation}
The displacement of the center--of--mass and the relative modes of
vibration can be done, independently of each other, by using the
excitation mechanism applied in Ref. \cite{wine1}. Our problem, then,
is to find a way to measure the displaced vibrational population,
$\Pi_{n_c,n_{r}}(-\alpha _{c},-\alpha _{r})$. For this purpose, we use
the same dispersive interaction scheme that led to the generation of
the $|\Phi^{\pm}\rangle$ Bell states. Consider that, after the
vibrational displacements, the state of the ions is described by
$\hat\rho(0)=\hat{\varrho}_v(-\alpha_c,-\alpha_r)\, \otimes$
$|\!\!\downarrow , \downarrow\rangle \langle \downarrow , \downarrow\!\!|$.
From Eqs.~(\ref{Heff}) and (\ref{evol}), the probability
$P_{\downarrow \downarrow}(\tau )$ of finding both ions in their lower
electronic states, after an interaction time $\tau $ with the laser
fields, is
\begin{eqnarray}
P_{\downarrow \downarrow }(\tau ) &=&\left\langle \downarrow , \downarrow 
\right| {\rm Tr}_{v}\left\{ e^{-\frac{i}{\hbar}\hat{H}_{\rm eff}\tau} \hat{\rho}(0)
e^{\frac{i}{\hbar}\hat{H}_{\rm eff}\tau}\right\} \left| \downarrow , \downarrow \right\rangle   \nonumber \\
&=&\sum_{n_c,n_{r}}\cos ^{2}\left( |\Omega _{n_cn_{r}}^{k}|\,\tau \right)
\Pi_{n_c,n_r}(-\alpha_c,-\alpha_r) \, ,  \label{prob}
\end{eqnarray}
where ${\rm Tr}_{v}$ means the partial tracing over the
vibrational degrees of freedom. From Eq.~(\ref{rabi}) one can see that, for 
$k\neq 0$, the dependence of the vibronic Rabi frequencies $\Omega
_{n_cn_{r}}^{k}$ on $n_c$ and $n_{r}$ discriminate efficiently the different two-mode
Fock states $\left| n_c,n_{r}\right\rangle $ (see Fig.~2). For this reason the
 vibrational populations $\Pi_{ n_c,n_{r}}(-\alpha_c, -\alpha_r) $ may be
easily extracted from $P_{\downarrow \downarrow }(\tau )$\cite{leibfried}.

\vspace*{-2.2cm}
\begin{figure}
\begin{center}
\centerline{\leavevmode \epsfxsize=9cm\epsfbox{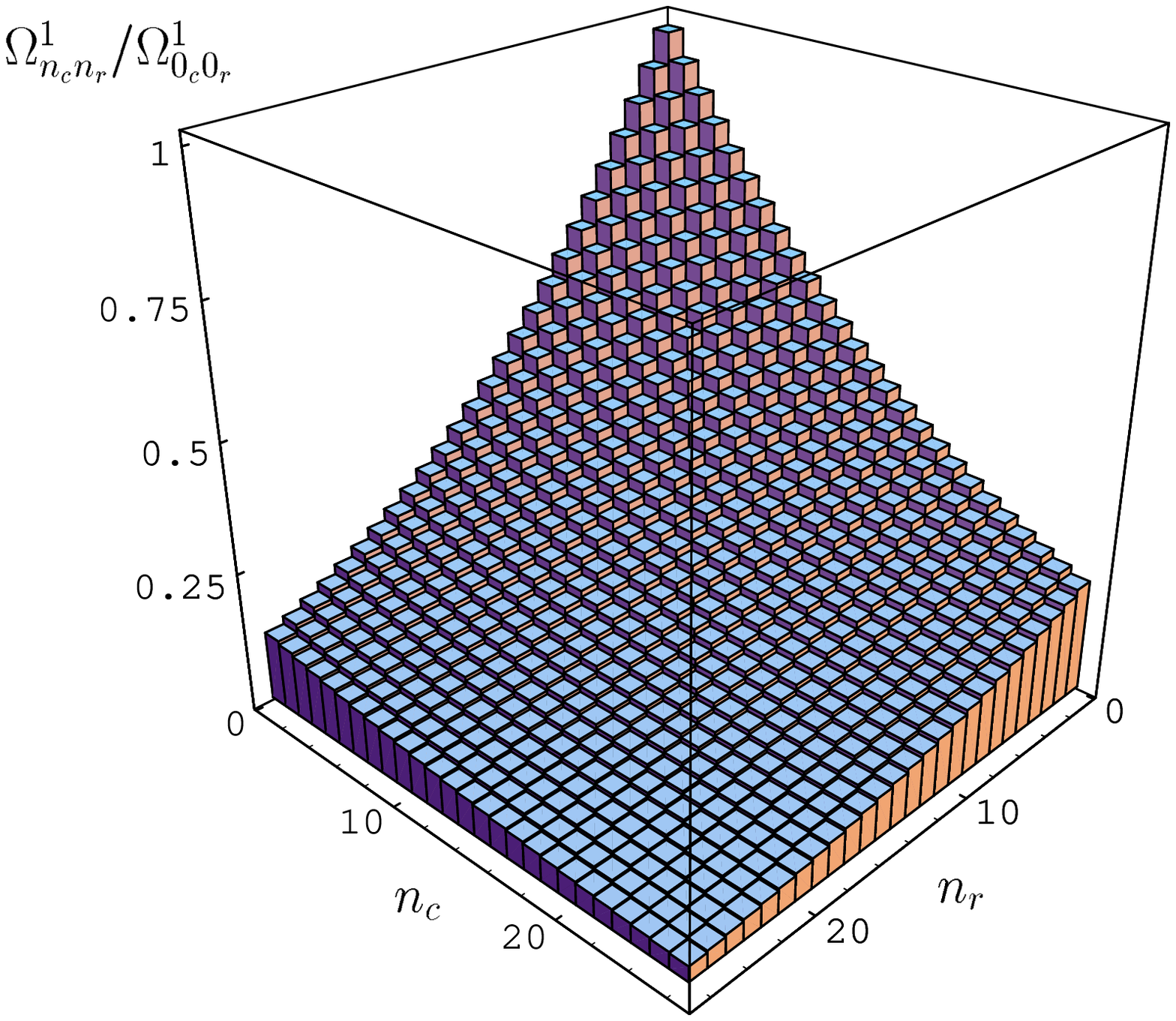}}

\vspace*{-3.0cm}
\caption{ Scaled Rabi frequency $\Omega _{n_{c}n_{r}}^{k}$ plotted as a
  function of $n_{c}$ and $n_{r}$ for $k=1$ and $\eta=0.23$. We stress
  the fact that we have a different positive Rabi frequency for each
  different pair $(0,0)<(n_{c},n_{r})<(25,25)$. }
\label{fig2}
\end{center}
\end{figure}

We can measure $P_{\downarrow \downarrow }(\tau )$ by monitoring the
fluorescence produced in driving the auxiliary cycling transition
$|\downarrow \rangle \leftrightarrow |d\rangle $ of the ions with
resonant laser light, once the interaction leading to
Eq.~(\ref{prob}) has been turned off (see Fig.~1). The presence of
fluorescence assures us that the ions are in the joint electronic state
$|\!\downarrow ,\downarrow \rangle $, and its absence indicates the
occupation of the state $|\!\uparrow ,\uparrow \rangle $. This is only
the case because the states $|\!\downarrow,\uparrow\rangle$ and
$|\!\uparrow,\downarrow\rangle$ are not populated before the cycling
transition is driven. That is, the use of the dispersive interaction 
 (Eq.~(\ref{Heff})) opens the possibility of measuring $
 P_{\downarrow \downarrow }(\tau )$ in as simple and highly efficient
 manner as that used to measure the corresponding quantity in a single
 ion~\cite {wine1}.

In summary, we have presented a method that allows the deterministic
generation of all the electronic Bell states of two trapped ions.
This is achieved by combining a purely dispersive with a resonant
laser excitation of vibronic transitions of the two ions.  It is
important to stress that, in the Lamb--Dicke limit, the generation of
the Bell states does not require previous cooling of the ions to their
motional ground state. Moreover, we have shown that the two-photon
dispersive interaction discussed in this Rapid Communication is highly adequate for
the implementation of a procedure to completely determine the motional
state of the ions.  In contrast to other methods presented up to now,
our proposal does not require the differential laser addressing of the
individual ions and, therefore, may be easily implemented with 
present available techniques.

This work was partially supported by the Conselho Nacional de
Desenvolvimento Cient{\'\i}fico e Tecnol\'ogico (CNPq) and the
Programa de Apoio a N\'ucleos de Excel\^encia (PRONEX).

\end{document}